\documentclass [12pt,a4paper]{article}
\usepackage {graphicx}
\begin{document}

\begin{center}

 \

{\large {\bf Equilibrium distribution functions}}

\vspace{0.2cm}

{\large {\bf of a homogeneous multicomponent gas mixture;}}

\vspace{0.2cm}

{\large{\bf inequality of partial temperatures}}

\vspace{0.5cm}

{\sc Yurii M.Loskutov}

\vspace{0.3cm}

{\sf Physics Faculty, Moscow State University, Moscow 119899 Russia}

\vspace{1.0cm}

{\bf Abstract}

\end{center}

It is shown that the partial temperatures of a homogeneous multicomponent gas
mixture in the thermodynamical equilibrium cannot be equal to each other.
New general solutions for equilibrium distribution functions of the multicomponent
mixture are found. Parameters (including partial temperatures) involved in these
solutions are determined by means of physical conditions which follow from a notion
of the stable thermodynamical equilibrium. The found relations of the partial
temperatures to the mean mixture temperature are dependent on the molecular
weights and concentrations of the components. Some possible consequences
of the developed approach are discussed.

\newpage\

\baselineskip=20pt

\

\section{General remarks}

There is a considerable body of work on the study of the distribution
functions and kinetic equations (see, e.g., [1]--[12] -- this list is quite imcomplete).
However up to the present a certain question is overlooked. As shown below, this
question is of fundamental importance because it qualitatively changes traditional
view on the thermodynamical equilibrium state.

Imagine that we observe collisions of molecules in a mixture of free gases
(i.e. gases that are subject to no external field) in the state of
thermodynamical equilibrium. Then the first thing we should notice is that for
every time interval
$\Delta t_0$, which equals the mean free time, the number of collisions
of particles with velocities ${\bf v}_1\cdot{\bf v}_2<0$ (I-type collisions)
will exceed\footnote{In principle, this statement holds not only for binary
collisions.} the number of collisions of particles with velocities
${\bf v}_1\cdot{\bf v}_2>0$ (II-type collisions). This is a direct consequence of
the well-known inequality of probabilities of I-type and II-type collisions.
If the mixture
consists of two components then the above considerations can be applied to
collisions of similar (i.e., of equal masses) molecules as well as to
collisions of different ones. Therefore the case of equal mean kinetic
energies of molecules of "light" and "heavy" components cannot correspond to
the thermodynamical equilibrium state (in equilibrium $U={\rm const}$ where
$U$ is the internal energy). Indeed, it is well known from
classical mechanics that in I-type collisions of particles with equal kinetic
energies light particles gain energy while heavy ones lose it. In the case of
II-type collisions the picture is the opposite. So, as the probabilities of
I-type collisions exceed the probabilities of II-type ones, the equality of
mean kinetic energies of light and heavy molecules cannot be maintained:
the light component will increase its energy to some extent while the heavy
component energy will decrease respectively. The energy transfer from
the heavy particles to the light ones will stop when the mean energy of
molecules of the "light" component exceeds at a certain value the mean energy
of the "heavy" component. Now in every time interval $\Delta t_0$ the mean
collision-induced change $\overline{\Delta T_\alpha}$ of kinetic energy of
molecules of the two components ($\alpha=1$ and $\alpha=2$) becomes zero.
Obviously it is this state which should be called equilibrium.

Notice that if the mean energy change is determined with respect to a
time interval $dt$ smoothed over many collisions we wouldn't be able to
follow the process of energy redistribution at a "discrete" level of individual
collisions. In other words, the condition $\overline{dT_\alpha}/dt=0$
generally would not allow to estimate the ratio of the mean
kinetic energies of molecules of one and the other kind at the equilibrium
state. This means that the condition $\overline{dT_\alpha}/dt=0$ corresponds
in general to a whole set of solutions for distribution functions (see
Section 2 below). Unlike this, the condition $\overline{\Delta T_\alpha}=0$
would imply the state of thermodynamical equilibrium with no energy
redistribution between the components, so this condition corresponds to
well-defined distribution functions (see Section 2 below). It is these
functions that lead to an inequality of mean kinetic energies
$\overline{T_\alpha}$ of different kind molecules at the equilibrium state.

This conclusion means that the representation of binary distribution function
as a product of the one-particle (Maxwell-type) functions would contradict the
necessary condition of absence of energy redistribution between
the components at the equilibrium state. To show this, consider the
BBGKI-type equations restricted to the case of binary collisions:
$$
   \frac{\partial f_i}
        {\partial t}
   +
   ({\bf v}_i\cdot\nabla_i)f_i
   =
   \sum_k n_k
          \int \frac{\partial\Phi_{ik}(|{\bf r}_k-{\bf r}_i|)}
                    {\partial{\bf r}_i}
               \frac{\partial F_{ik}}
                    {\partial{\bf p}_i}\,
               d^3x_k\,
               d^3p_k
   ,
   \eqno{(1)}
$$
$$
   \frac{\partial F_{ik}}
        {\partial t}
   +
   ({\bf v}_i\cdot\nabla_i)F_{ik}
   +
   ({\bf v}_k\cdot\nabla_k)F_{ik}
   -
$$
$$
   -
   \frac{\partial\Phi_{ik}(|{\bf r}_k-{\bf r}_i|)}
        {\partial{\bf r}_i}
   \frac{\partial F_{ik}}
        {\partial{\bf p}_i}
   -
   \frac{\partial\Phi_{ik}(|{\bf r}_k-{\bf r}_i|)}
        {\partial{\bf r}_k}
   \frac{\partial F_{ik}}
        {\partial{\bf p}_k}
   =
   0
   ,
   \eqno{(2)}
$$
where all the notations are standard. In view of the conditions imposed,
the derivative $\partial F_{ik}/\partial t$ in equations (2) can be
ignored. Then after an integration over $x_k, p_k$ one obtains
$$
   \int \frac{\partial\Phi_{ik}}
             {\partial{\bf r}_i}
        \frac{\partial F_{ik}}
             {\partial{\bf p}_{i}}\,
        d^3x_k\,
        d^3p_k
   =
   \int (
           {\bf v}_k
           -
           {\bf v}_i
        )
        \frac{\partial F_{ik}}
             {\partial{\bf r}_{ik}}\,
        d^3x_{ik}\,
        d^3p_k
   ,
$$
where ${\bf r}_{ik}={\bf r}_k-{\bf r}_i$. Applying now the common (see, e.g.,
[9, 10, 12]) procedure of integration with respect to $x_{ik}$ and
letting\footnote{The notation (3) merely symbolises optionality of
factorizing binary functions in single ones in transition to collision
integrals with the use of unitary operators which "move aside" the events at
$|{\bf r}_{ik}|\to\infty$ to the states before and after scattering.}
$$
   \left.
      F_{ik}
   \right|_{|{\bf r}_{ik}|\to\infty}
   \to
   F_{ik}(t,{\bf p}_i,{\bf p}_k)
   ,
   \eqno{(3)}
$$
one puts the equations (1) in the form
$$
   \frac{\partial f_1}
        {\partial t}
   =
   n_1
   \int d^3\tilde{v}_1\,
        d\sigma_{11}\,
        u_{11}
        \bigl[
           F_{11}^\prime({\bf v}_1^\prime,\tilde{\bf v}_1^\prime,t)
           -
           F_{11}({\bf v}_1,\tilde{\bf v}_1,t)
        \bigr]
   +
$$
$$
   +
   n_2
   \int d^3v_2\,
        d\sigma_{12}\,
        u_{12}
        \bigl[
           F_{12}^\prime({\bf v}_1^\prime,{\bf v}_2^\prime,t)
           -
           F_{12}({\bf v}_1,{\bf v}_2,t)
        \bigr]
   ,
$$
$$
   \eqno{\smash{(4)}}
$$
$$
   \frac{\partial f_2}
        {\partial t}
   =
   n_2
   \int d^3\tilde{v}_2\,
        d\sigma_{22}\,
        u_{22}
        \bigl[
           F^\prime_{22}({\bf v}^\prime_2,\tilde{\bf v}^\prime_2,t)
           -
           F_{22}({\bf v}_2,\tilde{\bf v}_2,t)
        \bigr]
   +
$$
$$
   +
   n_1
   \int d^3v_1\,
        d\sigma_{21}\,
        u_{21}
        \bigl[
           F^\prime_{21}({\bf v}^\prime_2,{\bf v}^\prime_1,t)
           -
           F_{21}({\bf v}_2,{\bf v}_1,t)
        \bigr]
   .
$$
Here $u_{11}\equiv|{\bf v}_1-\tilde{\bf v}_1|$,
$u_{22}\equiv|{\bf v}_2-\tilde{\bf v}_2|$,
$u_{12}=u_{21}\equiv|{\bf v}_2-{\bf v}_1|$, $d\sigma_{ik}$ are
the differential scattering cross-sections of $(i,\,k)$ molecules, and
the primed symbols correspond to the state after scattering; besides that, a
transition from integrals over momentum space to integrals over a space of
velocities was made for later convenience.

In carrying out the limit (3) a hypothesis of correlation moderation principle
is usually used, that is the limiting function $F_{ik}$ is reduced to
a product of the one-particle functions
$$
   F_{ik}(t,{\bf p}_i,{\bf p}_k)
   =
   f_i(t,{\bf p}_i)\,
   f_k(t,{\bf p}_k).
   \eqno{(3{\rm a})}
$$
Thus the system (4) becomes closed with respect to single functions, its
solutions in the case of thermodynamical equilibrium being Maxwell
distributions
$$
   f_i
   =
   A_i
   \exp(
          -m_i
           {\bf v}_i^2
           /
           2
           \theta
       )
   \eqno{(5)}
$$
with one parameter $\theta$ which determines the temperature of the mixture.

In the equations (4) there appear probabilities
$d{\rm w}_{ik}=n_ku_{ik}d\sigma_{ik}$ of collision and scattering of
particles per unit time. If one replaces these probabilities by
probabilities of single collision between particles with velocities ${\bf v}_i$,
${\bf v}_k$ per unit volume
$$
   dw_{ik}
   \equiv
   n_{ik}
   d{\rm w}_{ik}
   /
   n_k
   \sigma_{ik}
   \int u_{ik}\,
        d\Omega
   =
$$
$$
   =
   \frac{3n_{ik}}
        {2\pi}
   \frac{d\sigma_{ik}}
        {\sigma_{ik}}
   \frac{v_i v_k |{\bf v}_k - {\bf v}_i|}
        {|v_k + v_i|^3 - |v_k - v_i|^3}
   \eqno{(6)}
$$
following from the expression for $d{\rm w}_{ik}$ ($n_{ik}$ being a number of
such collisions per unit volume), then the corresponding collision integrals
in (4) will give the changes of one-particle functions under single collisions per
unit volume. Replacement of $d{\rm w}_{ik}$ by $dw_{ik}$ in (4) together with
substitution of $\Delta f_i$ instead of $\partial f_i/\partial t$ will have
no effect on the distribution functions in the thermodynamical
equilibrium state. As for conditions of absence of energy redistribution between
the components, they will acquire the form\footnote{To find the mean
collision-induced change $\overline{\Delta T_\alpha}$,
we should take the difference
$\Delta T_\alpha\equiv T^\prime_\alpha- T_\alpha$ multiply it by the
probability of collision and scattering $dw_{ik}$ and after this
to sum over all finite states and average (by means of $F_2$)
over all initial states. The second condition $\overline{\Delta T_2}=0$
will be a consequence of the first one in the case of a two-component mixture.}
$$
   \overline{\Delta T_1}
   =
   \int d^3v_1\,
        d^3\tilde v_1\,
        dw_{11}\,
        F_{11}\Delta T_1
   +
   \int d^3v_1\,
        d^3v_2\,
        dw_{12}\,
        F_{12}\Delta T_1
   =
   0
   .
   \eqno{(7)}
$$

The increment $\Delta T_1\equiv T_1^\prime-T_1$ caused by a collision of
molecules with masses $m_1$ and $m_2$ is determined by
$$
   T_1^\prime
   -
   T_1
   =
   \mu
   ({\bf u}\cdot{\bf v})
   -
   \mu
   ({\bf u}^\prime\cdot{\bf v})
   ,
   \eqno{(8)}
$$
where $\mu\equiv m_1m_2/m$, $m\equiv m_1+m_2$,
${\bf u}\equiv{\bf v}_2-{\bf v}_1$,
${\bf u}^\prime\equiv{\bf v}_2^\prime-{\bf v}_1^\prime$,
${\bf v}\equiv(m_1{\bf v}_1+m_2{\bf v}_2)/m$. Substituting in (7) the expressions
(8) and (3a) with account for (5) it is easy to see that the first integral in
(7) vanishes while the second, which is connected with collisions of
molecules of different masses, remains nonzero. This means that
the solutions of the form (3a) based on correlation moderation principle do not satisfy
the condition of absence of energy redistribution between light and heavy
components. Therefore solutions (3a) with the only parameter $\theta$ should
be replaced by more general solutions depending not only on the energy
but also on other integrals of motion, that is solutions with a large number
of parameters.

As such, a possibility of introducing of various integrals of motion into
solutions of the kinetic equations doesn't contradict anything and is very
natural because a Liouville equation for the distribution function of
a system of $N$ particles can in principle include $6N$ integrals of motion.
Still the question remains how does this agree with the Gibbs distribution
which is deduced from rather general considerations? It should be noted here
that the Gibbs canonical distribution is obtained under the assumption
(guaranteed for systems of a very large number of particles) that boundaries
contribute only an ignorable part of interaction energy of neighbouring
regions. It will be shown in the following sections that the general solution
for equilibrium distribution functions involves terms much smaller than
the boundary ones inevitably arising in the traditional derivation of the Gibbs
distribution. If one ignores them\footnote{Strictly speaking, they are
neglected in deriving the Gibbs distribution as well, because the densities
introduced there are searched for as functions of a energy integral,
other integrals of motion (e.g., momentum) being ignored.}
the general solution will transform into the Gibbs distribution with all
the corresponding implications. On the contrary, if one takes them into
account in transition to the single and binary distribution functions, these small
terms will give a contribution comparable with that of the energy integral.

Thus the above considerations imply that in general the equilibrium
distribution functions should depend on several integrals of motion, which
involve several free parameters. These parameters are determined by
the following physical conditions:
\begin{enumerate}
   \item The one-particle equilibrium distribution functions of molecules of some
         kind derived from different binary or many-particle distribution
         functions should be identical.
   \item In order for the state of thermodynamical equilibrium to be stable one
         should require that the energy density be minimal in this
         state.\footnote{In the problem in question this condition is
         equivalent to that of minimality of free energy, as the state of
         thermodynamical equilibrium with constant number of particles and
         constant entropy $S$ and temperature $\theta$, characterizing
         the whole of gas mixture, is considered.}
   \item One should require that the average changes of kinetic energy of
         molecules per unit volume caused by one-particle collisions be equal to
         zero to guarantee an absence of energy redistribution between
         the components.
\end{enumerate}
It will be shown below that these conditions lead to uniquely determined
expressions for equilibrium distribution functions.

\section{General solutions of kinetic equations for distribution functions of
a homogeneous two-com\-po\-nent mixture of free gases
in thermodynamical equilibrium}

Let the system consist of a homogeneous mixture of $N_1$ molecules with mass
$m_1$ and $N_2$ molecules with mass $m_2$. Then, as is well known,
the equations (1), (2) will be satisfied at the state of thermodynamical
equilibrium by arbitrary functions of two variables (separated by semicolon)
$$
   F_{ik}
   =
   F_{ik}\left(
            \mu_{ik}
            \left(
               \frac{\stackrel{\smash{\scriptscriptstyle(i)}}{{\bf p}}}
                    {m_i}
               -
               \frac{\stackrel{\smash{\scriptscriptstyle(k)}}{{\bf p}}}
               {m_k}
            \right)^2
            ;\;
            \stackrel{\smash{\scriptscriptstyle(i)}}{{\bf p}}
            +
            \stackrel{\smash{\scriptscriptstyle(k)}}{{\bf p}}
         \right)
   .
   \eqno{(9)}
$$
In order for one-particle distribution functions obtained from the binary ones to
have the form of Maxwell-type distributions the two-parameter solutions (9) should
be an exponential of quadratic combinations of momenta. Thus,
the general solution $F_{12}$ should have the form
$$
   F_{12}
   =
   A_{12}
   \exp\left\{
          -C_1
          \mu_{12}
          \left(
             \frac{\stackrel{\smash{\scriptscriptstyle(1)}}{{\bf p}}_1}
                  {m_1}
             -
             \frac{\stackrel{\smash{\scriptscriptstyle(2)}}{{\bf p}}_1}
             {m_2}
          \right)^2
          -
          C_2
          \bigl(
             \stackrel{\smash{\scriptscriptstyle(1)}}
                      {{\bf p}}_1
             +
             \stackrel{\smash{\scriptscriptstyle(2)}}
                      {{\bf p}}_1
          \bigr)^2
       \right\},
$$
where $\stackrel{\smash{\scriptscriptstyle(1)}}{{\bf p}}_1$ and
$\stackrel{\smash{\scriptscriptstyle(2)}}{{\bf p}}_1$ denote momenta of
particles with masses $m_1$ and $m_2$, respectively. In what follows it will
be more convenient to introduce two parameters $\theta_{12}$ and $\lambda$
instead of two constants of integration $C_1$ and $C_2$, so that
$$
   C_1\equiv(1-\lambda)/2\theta_{12},\qquad
   C_2\equiv(1+\lambda)/2m\theta_{12}.
$$
Dealing with $F_{11}$ and $F_{22}$ in a similar way, we
obtain\footnote{In [13] the solutions for distribution functions of similar
molecules are taken with the parameter $\lambda_0=0$. Such solutions satisfy
the system of equations (1), (2) as well. It would be more consistent,
however, to consider $\lambda_0$ as the limit of $\lambda(\varepsilon)$ as
$\varepsilon\to 0$, where $\varepsilon\equiv(m_2-m_1)/m$.}
$$
   F_{11}
   =
   A_{11}
   \exp\left\{
          -\frac{1}
                {2\,(1 + \lambda_0)\,\theta_1 m_1}
          \left[
             \stackrel{\smash{\scriptscriptstyle(1)}}
                      {{\bf p}}_1^2
             +
             \stackrel{\smash{\scriptscriptstyle(1)}}
                      {{\bf p}}_2^2
             +
             \frac{\lambda_0}
                  {1 - \lambda_0}
             \bigl(
                \stackrel{\smash{\scriptscriptstyle(1)}}
                         {{\bf p}}_1
                +
                \stackrel{\smash{\scriptscriptstyle(1)}}
                         {{\bf p}}_2
             \bigr)^2
          \right]
       \right\}
   ,
$$
$$
   F_{22}
   =
   A_{22}
   \exp\left\{
          -\frac{1}
                {2\,(1+\lambda_0)\,\theta_2 m_2}
          \left[
             \stackrel{\smash{\scriptscriptstyle(2)}}
                      {{\bf p}}_1^2
             +
             \stackrel{\smash{\scriptscriptstyle(2)}}
                      {{\bf p}}_2^2
             +
             \frac{\lambda_0}
                  {1-\lambda_0}
             \bigl(
                \stackrel{\smash{\scriptscriptstyle(2)}}
                         {{\bf p}}_1
                +
                \stackrel{\smash{\scriptscriptstyle(2)}}
                         {{\bf p}}_2
             \bigr)^2
          \right]
       \right\}
   ,
   \eqno{(10)}
$$
$$
   F_{12}
   =
   A_{12}
   \exp\left\{
          -\frac{1-\lambda}
                {2\theta_{12} m}
          \left[
             \frac{m}
                  {m_1}
             \stackrel{\smash{\scriptscriptstyle(1)}}
                      {{\bf p}}_1^2
             +
             \frac{m}
                  {m_2}
             \stackrel{\smash{\scriptscriptstyle(2)}}
                      {{\bf p}}_1^2
             +
             \frac{2\lambda}
                  {1-\lambda}
             \bigl(
                \stackrel{\smash{\scriptscriptstyle(1)}}
                         {{\bf p}}_1
                +
                \stackrel{\smash{\scriptscriptstyle(2)}}
                         {{\bf p}}_1
             \bigr)^2
          \right]
       \right\}
   .
$$
Here the superscripts (1), (2) of the momenta ${\bf p}_i$ refer to
the molecular species $(m_1,m_2)$, $\lambda=\lambda(\varepsilon),\
\varepsilon\equiv (m_2 - m_1)/(m_2 + m_1)$, and
$\lambda_0\equiv\lambda(0)$. The parameter $\lambda$ describes the
degree of correlation of particle momenta in equilibrium.
It is easy to see that the conditions $\overline{dT_\alpha}/dt=0$
corresponding to (7) under substitution of $d{\rm w}_{ik}$ instead of
$dw_{ik}$ are satisfied by a family of equilibrium solutions (10) with
arbitrary $\lambda_0$ and $\lambda$, as was stated in Section 1. If one
assumes $\lambda=\lambda_0=0$ in (10), functions $F_{ik}$ will factorize and
partial temperatures $\theta_1$ and $\theta_2$ will become equal
to each other by virtue of the condition 1 of Section 1.
This is in contradiction to the condition 3 of Section 1. Hence
particular solutions with $\lambda(\varepsilon)=0$ cannot correspond to
thermodynamical equilibrium, i.e. more general solutions with
$\lambda(\varepsilon)\ne 0$ should be taken as equilibrium ones.

The solutions (10) also arise from a many-particle distribution function $F$
satisfying the stationarity condition on every interval $\Delta t_0$.
It has the form
$$
   F
   =
   A
   \exp\left\{
          -\frac{1}
                {2\,(1+\lambda_0)\,\theta_1 m_1}
          \left[
             \sum_{i_1=1}^{N_1^{(11)}}\stackrel{\smash{\scriptscriptstyle(1)}}
                                               {{\bf p}}_{i_1}^2
             +
             \frac{\lambda_0 {\bf P}_{(11)}^2}
                  {1-(N_1^{(11)}-1)\,\lambda_0}
          \right]
          -
       \right.
$$
$$
          -
          \frac{1}
               {2\,(1+\lambda_0)\,\theta_2 m_2}
          \left[
             \sum_{i_2=1}^{N_2^{(22)}}\stackrel{\smash{\scriptscriptstyle(2)}}
                                               {{\bf p}}_{i_2}^2
             +
             \frac{\lambda_0 {\bf P}_{(22)}^2}
                  {1-(N_2^{(22)}-1)\,\lambda_0}
          \right]
          -
$$
$$
          -
          \frac{1-\lambda}
               {2\theta_{12} m}
          \left[
             \frac{m}
                  {m_1}
             \sum_{i_3=1}^{N_1^{(12)}}\stackrel{\smash{\scriptscriptstyle(1)}}
                                               {{\bf p}}_{i_3}^2
             +
             \frac{m}
                  {m_2}
             \sum_{i_3=1}^{N_2^{(12)}}\stackrel{\smash{\scriptscriptstyle(2)}}
                                               {{\bf p}}_{i_3}^2
             +
          \right.
$$
$$
       \left.
          \left.
             +
             \frac{2\lambda {\bf P}_{(12)}^2}
                  {
                     1
                     -
                     \left(
                        N_1^{(12)}
                        +
                        N_2^{(12)}
                        -
                        1
                     \right)\,
                     \lambda
                     +
                     \left(
                        N_1^{(12)}
                        -
                        N_2^{(12)}
                     \right)\,
                     \varepsilon
                     \lambda
                  }
          \right]
       \right\}
   ,
   \eqno{(11)}
$$
where
$$
   {\bf P}_{(11)}
   \equiv
   \sum_{i_1=1}^{N_1^{(11)}}\stackrel{\smash{\scriptscriptstyle(1)}}
                                     {{\bf p}}_{i_1}
   ,
   {\bf P}_{(22)}
   \equiv
   \sum_{i_2=1}^{N_2^{(22)}}\stackrel{\smash{\scriptscriptstyle(2)}}
                                     {{\bf p}}_{i_2}
   ,
   {\bf P}_{(12)}
   \equiv
   \sum_{i_3=1}^{N_1^{(12)}}\stackrel{\smash{\scriptscriptstyle(1)}}
                                     {{\bf p}}_{i_3}
   +
   \sum_{i_3=1}^{N_2^{(12)}}\stackrel{\smash{\scriptscriptstyle(2)}}
                                     {{\bf p}}_{i_3}
   ,
$$
indices $i_1$, $i_2$, and $i_3$ label the molecules corresponding to
collisions $(m_1,\,m_1)$, $(m_2,\,m_2)$, and $(m_1,\,m_2)$,
while $N_1^{(11)}$, $N_2^{(22)}$ and $N_1^{(12)}$, $N_2^{(12)}$
are the numbers of molecules of the two kinds in the associated groups,
the relations $N_1^{(11)}+N_1^{(12)}=N_1$, $N_2^{(22)}+N_2^{(12)}=N_2$ being
satisfied.

In the case of a large number of molecules the terms in (11) with
$N_\alpha^{(\alpha\beta)}$ in the denominator will be small compared to
the others, since, due to the chaotic nature of molecular
motion\footnote{The reference frame is chosen so that the
mean momentum of molecules of  the mixture is equal to zero.},
the momenta ${\bf P}_{(\alpha\beta)}$ cannot
exceed the momenta $\stackrel{\smash{\scriptscriptstyle(\alpha)}}{\bf p}$ by
a significant amount. If one neglects these terms then by virtue of relations
$$
   \frac{1}
        {(1+\lambda_0)\,\theta_1}
   =
   \frac{1-\lambda}
        {\theta_{12}}
   =
   \frac{1}
        {(1+\lambda_0)\theta_2}
   =
   \frac{1}
        {\theta_0},
$$
following from the indistinguishability of molecules of the same kind,
one obtains the Gibbs distribution with the unique temperature
$\theta_0$ and all corresponding
implications for binary and single distribution functions. Unlike this, if one
takes these terms into account when deriving single and binary distribution
functions then their final contribution will turn out to be comparable with
that of other terms. Indeed, it can be easily verified that each integration over
one of the momenta $\stackrel{\smash{\scriptscriptstyle(\alpha)}}{\bf p}$ is
equivalent to decreasing of the corresponding $N_\alpha^{(\alpha\beta)}$ in
(11) by unit. Therefore the solutions (10) correspond to the solutions (11) with
$N_1^{(11)}=2$, $N_2^{(22)}=2$, and $N_1^{(12)}=N_2^{(12)}=1$, the other
$N_\alpha^{(\alpha\beta)}$ being equal to zero.

We can find solution (10) by another way.
Let us introduce a new value $\tilde H\equiv \log F_{12}$ and find its mean
change (but not the change of the mean) in a unit of time which takes
place by collision and scattering. To make this it is necessary to take
the difference in $\tilde H$ {\it before} and {\it after} collision
(i.e. ${\tilde H}^\prime -\tilde H$) multiply this difference by the
probability of collision and scattering in a unit time (i.e. by
$n_{12}u_{12}d\sigma_{12}$) and after this to integrate over all
finite states and average over initial (i.e. by means
of $F_{12}$ {\it before} scattering) states. Then
$$
\begin{array}{c}
\displaystyle
{\overline {\frac{d\tilde H}{dt}}}=
n_{12}\int(\log F^\prime_{12}-\log F_{12})F_{12}d\sigma_{12}
u_{12}d^3p_1d^3p_2d^3x_1d^3x_2= \\
 \\
\displaystyle
=-\frac{n_{12}}{2}\int(F^\prime_{12}-F_{12})
\log \frac{F^\prime_{12}}{F_{12}}d\sigma_{12}
u_{12}d^3p_1d^3p_2d^3x_1d^3x_2 \ \ .
\end{array}
$$
From this expression one can easily find that
$$
{\overline {\frac{d\tilde H}{dt}}}\leq 0 \ .
$$
The equality is achieved if $\log F^\prime_{12}=\log F_{12}$. Thus, $\log F_2$
is the additive integrals of the motion of two bodies that is described by (10).

In accordance with the condition 1 of Section 1, imposing a requirement of
the identity of one-particle distribution functions derived from $F_{ii}$ and
$F_{ik}$, that is demanding that
$$
   f_1
   =
   \int F_{11}\,
        d^3\stackrel{\smash{\scriptscriptstyle(1)}}
                    {{\bf p}}_2
   =
   \int F_{12}\,
        d^3\stackrel{\smash{\scriptscriptstyle(2)}}
                    {{\bf p}}_1
   =
   A_1
   \exp\left\{
          -\stackrel{\smash{\scriptscriptstyle(1)}}
                    {{\bf p}}_1^2
          /
          2
          m_1
          \theta_1
       \right\}
   ,
$$
$$
   f_2
   =
   \int F_{22}\,
        d^3\stackrel{\smash{\scriptscriptstyle(2)}}
                    {{\bf p}}_2
   =
   \int F_{12}\,
        d^3\stackrel{\smash{\scriptscriptstyle(1)}}
                    {{\bf p}}_1
   =
   A_2
   \exp\left\{
          -\stackrel{\smash{\scriptscriptstyle(2)}}
                    {{\bf p}}_1^2
                    /
                    2
                    m_2
                    \theta_2
       \right\}
   ,
$$
one obtains the relations between the parameters\footnote{These relations
could also be obtained from (11) if one lets $N_1^{(11)}=1$ or $N_1^{(12)}=1$
and $N_2^{(22)}=1$ or $N_2^{(12)}=1$, the other $N_\alpha^{(\alpha\beta)}$ being
equal to zero.}
$$
   \theta_1
   =
   \theta_{12}
   \frac{1+\varepsilon\lambda}
        {1-\lambda^2},\qquad
   \theta_2
   =
   \theta_{12}
   \frac{1-\varepsilon\lambda}
        {1-\lambda^2}
   .
   \eqno{(12)}
$$
To establish a relation between the temperatures $\theta_1$, $\theta_2$, and
$\theta_{12}$ and the mean temperature of the mixture $\theta$ we make use
the condition of minimality of the energy density $E$ of the system in
thermodynamical equilibrium formulated at the end of Section 1:
$\partial E(\lambda)/\partial\lambda=0$, where
$E(\lambda)=n_1\theta_1+n_2\theta_2$, $\theta_1$ and $\theta_2$ being
determined by (12). This implies
$$
   \theta_{12}
   =
   C
   \frac{1-\lambda^2}
        {n+(n_1-n_2)\,\varepsilon\lambda}
   ,
$$
where $n=n_1+n_2$. The constant $C=n_1\theta_1+n_2\theta_2$, which equals to
the mean kinetic energy density of the mixture, may be set also as $n\theta$.
Thus one finds
$$
   \theta_{12}
   =
   \frac{(1-\lambda^2)\,n\theta}
        {n+(n_1-n_2)\,\varepsilon\lambda}
   ,
$$
$$
   \eqno{\smash{(13)}}
$$
$$
   \theta_1
   =
   \frac{(1+\varepsilon\lambda)\,n\theta}
        {n+(n_1-n_2)\,\varepsilon\lambda}
   ,
   \theta_2
   =
   \frac{(1-\varepsilon\lambda)\,n\theta}
        {n+(n_1-n_2)\,\varepsilon\lambda}
   .
$$
We can obtain expressions (13) from (12) if we {\it require}
that the mean temperature of the mixture ($\theta$)
will be determined by the formula
$\theta=(n_1/n)\theta_1 +(n_2/n)\theta_2$
that with the physical point of view is quite clear.

A remarkable thing here is the dependence of partial equilibrium temperatures
$\theta_1$ and $\theta_2$ on molecular weights and concentrations of
the components. If due to some reasons the molecular mixture splits into
a set of subsystems with an substantial domination of molecules of some sort
in each of them (with $n_2\gg n_1$ in some subsystems and $n_1\gg n_2$ in
others) then the partial temperatures $\theta_1$ and $\theta_2$ of subsystems
will practically become equal --- see (13). This effect is due to the fact that
collisions of molecules of different types which account for inequality of
$\theta_1$ and $\theta_2$ will occur only in the vicinity of boundaries of
subsystems. In view of the above the heat transfer from
a cold body to a hot one becomes impossible.

Finally, we establish a relation between $\lambda$ and $\varepsilon$ making
use of the third condition formulated at the end of Section 1, namely that of
absence of the energy redistribution between molecules of the mixture under
collisions in thermodynamical equilibrium. Taking into account in (7)
that the integral containing $F_{11}$ is identically zero, one obtains
the equation
$$
   \int d^3v_1\,
        d^3v_2\,
        (
           T_1^\prime
           -
           T_1
        )
        F_{12}\,
        dw_{12}
   =
   0
   ,
   \eqno{(7{\rm a})}
$$
where $T_1^\prime-T_1$ is defined in (8). After integration over scattering
angles and one of the velocities it can be cast into
the form\footnote{The proof of equality of mean kinetic energies of different
molecules suggested in [14] cannot be true because it is
overlooked there that the change $\Delta T_1$ is not a kinetic value but is
determined instead by a dynamical collision process which has a stochastic
character not taken into account in [14].}
$$
   \int\limits_{-\infty}^\infty
       dx\,
       \frac{x^2|1+x|^3}
            {|1+x|^3-|1-x|^3}\,
       \int\limits_0^1
           dy\,
           y^2
           \frac{x^2-1+\varepsilon y^2(1+x)^2}
                 {(x^2-2bxy^2+a^2)^4}
   =
   0
   ,
   \eqno{(14)}
$$
where
$$
   a^2
   \equiv
   (1-\varepsilon)
   (1-\varepsilon\lambda)
   /
   (1+\varepsilon)
   (1+\varepsilon\lambda),\qquad
   b
   \equiv
   \lambda
   (1-\varepsilon)
   /
   (1+\varepsilon\lambda)
   .
$$

Integration in (14) can be carried out only with respect to one
variable. Thus the dependence of $\lambda$ on $\varepsilon$ is to be
evaluated by numerical methods. The corresponding algorithm
turns out to be simpler if both integrations are left in (14). The numerical
values\footnote{In [13] there was an error in the numerical
algorithm found already after the publication. The corrected results were
published later [15].} of $\lambda$ corresponding to different values
of $\varepsilon$ are represented in Table 1 (1.000e-03 means
$1.000\cdot 10^{-3}$ and so forth). Figure 1 shows the dependence
graphically. As indicated there, the value of $\lambda$ is positive for all
$\varepsilon$ and tends to zero as $\varepsilon\to 1$ which is clear from
qualitative considerations: as $\varepsilon\to 1$ the probabilities of I-type
and II-type collisions tend to be equal together with $\theta_1$ and
$\theta_2$. Shown in figure 2 is the dependence of $\varepsilon\lambda$ on
$\varepsilon$ corresponding to the ratio of partial equilibrium temperatures
$\theta_1$ and $\theta_2$. It demonstrates that the maximal difference of
temperatures of light ($\theta_1$) and heavy ($\theta_2$) components is
achieved at molecular weight ratio $m_2/m_1\simeq 4.26$ and amounts to
15.5\% of $\theta_2$. Far from this point the difference
$\theta_1-\theta_2$ smoothly decreases and tends to zero as
$\varepsilon\to 0$ and $\varepsilon\to 1$.

\section{General solutions for equilibrium dis\-
tri\-bu\-tion fun\-c\-tions in the case
of multicomponent mixture}

For a multicomponent mixture which formed by molecules with masses
$m_1\le m_2\le\ldots\le m_s$, the binary distribution functions
in thermodynamical equilibrium expressed in terms of velocities will
be in conformity with (10):
$$
   F_{\alpha\alpha}
   =
   (
      1
      -
      \lambda_0^2
   )^{-3/2}
   \left(
      \frac{m_\alpha}
           {2\pi\theta_\alpha}
   \right)^3
   \exp\left\{
          -\frac{m_\alpha}
                {2\,(1-\lambda_0^2)\,\theta_\alpha}
          \left[
             \stackrel{\smash{\scriptscriptstyle(\alpha)}}
                      {{\bf v}}_1^2
             +
             \stackrel{\smash{\scriptscriptstyle(\alpha)}}
                      {{\bf v}}_2^2
             +
             2
             \lambda_0
             \stackrel{\smash{\scriptscriptstyle(\alpha)}}
                      {{\bf v}}_1
             \cdot
             \stackrel{\smash{\scriptscriptstyle(\alpha)}}
                      {{\bf v}_2}
          \right]
       \right\}
   ,
$$
$$
   F_{\alpha\beta}
   =
   \frac{[(1-\lambda_{\alpha\beta}^2)\,m_\alpha m_\beta]^{3/2}}
        {(2\pi\theta_{\alpha\beta})^3}
   \exp\Biggl\{
          -\frac{m_{\alpha\beta}}
                {4\theta_{\alpha\beta}}
          \Biggl[
             (
                1
                -
                \varepsilon_{\alpha\beta}
             )
             (
                1
                -
                \varepsilon_{\alpha\beta}
                \lambda_{\alpha\beta}
             )
             \stackrel{\smash{\scriptscriptstyle(\alpha)}}
                      {{\bf v}}_1^2
             +
$$
$$
             +
             (
                1
                +
                \varepsilon_{\alpha\beta}
             )
             (
                1
                +
                \varepsilon_{\alpha\beta}
                \lambda_{\alpha\beta}
             )
             \stackrel{\smash{\scriptscriptstyle(\beta)}}
                      {{\bf v}}_1^2
             +
             2
             \lambda_{\alpha\beta}
             (
                1
                -
                \varepsilon_{\alpha\beta}^2
             )
             \stackrel{\smash{\scriptscriptstyle(\alpha)}}
                      {{\bf v}}_1
             \cdot
             \stackrel{\smash{\scriptscriptstyle(\beta)}}
                      {{\bf v}}_1
          \Biggr]
       \Biggr\}
   .
   \eqno{(15)}
$$
Here $m_{\alpha\beta}\equiv m_\alpha+m_\beta$,
$\varepsilon_{\alpha\beta}=-\varepsilon_{\beta\alpha}\equiv(m_\beta-m_\alpha)/m_{\alpha\beta}$,
and the dependent quantities $m_{\alpha\beta}$ and $\varepsilon_{\alpha\beta}$
are related to the independent ones $m_{1\alpha}$ and $\varepsilon_{1\alpha}$ as
$$
   m_{\alpha\beta}
   =
   m_{1\alpha}
   +\varepsilon_{1\beta}
   m_{1\beta},\qquad
   \varepsilon_{\alpha\beta}
   =
   (
      \varepsilon_{1\beta}
      -
      \varepsilon_{1\alpha}
   )
   /
   (
      1
      -
      \varepsilon_{1\alpha}
      \varepsilon_{1\beta}
   )
   .
   \eqno{(16)}
$$
Functions (15) can be obtained from the many-particle distribution function
$$
   F
   =
   A
   \exp\Biggl\{
          -\frac{1}
                {2\,(1+\lambda_0)}
          \sum_{\alpha=1}^{s} \frac{1}
                                   {m_\alpha\theta_\alpha}
                              \Biggl[
                                 \sum_{i=1}^{N_\alpha^{(\alpha\alpha)}} \stackrel{\smash{\scriptscriptstyle(\alpha)}}
                                                                                 {{\bf p}}_i^2
                                 +
                                 \frac{\lambda_0{\bf P}_{(\alpha\alpha)}^2}
                                      {1-(N_\alpha^{(\alpha\alpha)}-1)\,\lambda_0}
                              \Biggr]
          -
$$
$$
          -
          \sum_{\beta>\alpha=1}^{s} \frac{1-\lambda_{\alpha\beta}}
                                         {2\theta_{\alpha\beta} m_{\alpha\beta}}
                                    \Biggl[
                                       \frac{m_{\alpha\beta}}
                                            {m_\alpha}
                                       \sum_{i=1}^{N_\alpha^{(\alpha\beta)}} \stackrel{\smash{\scriptscriptstyle(\alpha)}}
                                                                                      {{\bf p}}_i^2
                                       +
                                       \frac{m_{\alpha\beta}}
                                            {m_\beta}
                                       \sum_{i=1}^{N_\beta^{(\alpha\beta)}} \stackrel{\smash{\scriptscriptstyle(\beta)}}
                                                                                     {{\bf p}}_i^2
                                       +
$$
$$
                                       +
                                       \frac{
                                               2
                                               \lambda_{\alpha\beta}
                                               {\bf P}_{(\alpha\beta)}^2
                                            }
                                            {
                                               1
                                               -
                                               (
                                                  N_\alpha^{(\alpha\beta)}
                                                  +
                                                  N_\beta^{(\alpha\beta)}
                                                  -
                                                  1
                                               )\,
                                               \lambda_{\alpha\beta}
                                               +
                                               (
                                                  N_\alpha^{(\alpha\beta)}
                                                  -
                                                  N_\beta^{(\alpha\beta)}
                                               )\,
                                               \varepsilon_{\alpha\beta}
                                               \lambda_{\alpha\beta}
                                            }
                                    \Biggr]
       \Biggr\}
   ,
   \eqno{(17)}
$$
where
$$
   {\bf P}_{(\alpha\alpha)}
   \equiv
   \sum_{i=1}^{N_\alpha^{(\alpha\alpha)}} \stackrel{\smash{\scriptscriptstyle(\alpha)}}
                                                   {{\bf p}}_i
   ,\qquad
   {\bf P}_{(\alpha\beta)}
   \equiv
   \sum_{i=1}^{N_\alpha^{(\alpha\beta)}} \stackrel{\smash{\scriptscriptstyle(\alpha)}}
                                                  {{\bf p}_i}
   +
   \sum_{i=1}^{N_\beta^{(\alpha\beta)}} \stackrel{\smash{\scriptscriptstyle(\beta)}}
                                                 {{\bf p}_i}
   ,
$$
and $N_\alpha^{(\alpha\beta)}$ is a number of molecules with mass $m_\alpha$
taking part in $(m_\alpha,\,m_\beta)$-collisions. If in the case of large
numbers of molecules of different kinds one ignores in (17) the terms with
$N_\alpha^{(\alpha\beta)}$ in denominator then as in Section 2 one obtains
the Gibbs distribution with the single temperature common for all components.

Integration of (17) over one of the momenta
$\stackrel{\smash{\scriptscriptstyle(\alpha)}}{{\bf p}}_i$ is equivalent to
reducing the corresponding $N_\alpha^{(\alpha\beta)}$ in $F$ by unit, so (15)
corresponds to (17) with $N_\alpha^{(\alpha\alpha)}=2$ and
$N_\alpha^{(\alpha\beta)}=N_\beta^{(\alpha\beta)}=1$. The condition of
coincidence of one-particle functions obtained from different binary ones gives
the equalities
$$
   \theta_\alpha
   =
   \theta_{\alpha\beta}
   \frac{1+\varepsilon_{\alpha\beta}\lambda_{\alpha\beta}}
        {1-\lambda_{\alpha\beta}^2}
   ,\qquad
   \theta_\beta
   =
   \theta_{\alpha\beta}
   \frac{1-\varepsilon_{\alpha\beta}\lambda_{\alpha\beta}}
        {1-\lambda_{\alpha\beta}^2}
   .
$$
With the aid of these relations the dependent $\lambda_{\alpha\beta}$ are
expressed in terms of the independent ones:
$$
   \varepsilon_{\alpha\beta}
   \lambda_{\alpha\beta}
   =
   \frac{
           \varepsilon_{1\beta}
           \lambda_{1\beta}
           -
           \varepsilon_{1\alpha}
           \lambda_{1\alpha}
        }
        {
           1
           -
           \varepsilon_{1\alpha}
           \varepsilon_{1\beta}
           \lambda_{1\alpha}
           \lambda_{1\beta}
        }
   ,
$$
$$
   \eqno{\smash{(18)}}
$$
$$
   \left.
      \lambda_{\alpha\beta}
   \right|_{\alpha\neq\beta}
   =
   \left.
      \lambda_{\beta\alpha}
   \right|_{\alpha\neq\beta}
   =
   \frac{
           (
              \varepsilon_{1\beta}
              \lambda_{1\beta}
              -
              \varepsilon_{1\alpha}
              \lambda_{1\alpha}
           )
           (
              1
              -
              \varepsilon_{1\alpha}
              \varepsilon_{1\beta}
           )
        }
        {
           (
              1
              -
              \varepsilon_{1\alpha}
              \varepsilon_{1\beta}
              \lambda_{1\alpha}
              \lambda_{1\beta}
           )
           (
              \varepsilon_{1\beta}
              -
              \varepsilon_{1\alpha}
           )
        }
   .
$$
Making use of condition of minimality of energy density of the mixture we
obtain
$$
   \theta_{\alpha\beta}
   =
   n
   \theta
   \frac{
           (
              1
              -
              \lambda_{\alpha\beta}^2
           )
           (
              1
              -
              \varepsilon_{1\alpha}
              \varepsilon_{1\beta}
              \lambda_{1\alpha}
              \lambda_{1\beta}
           )
        }
        {
           (
              1
              +
              \varepsilon_{1\alpha}
              \lambda_{1\alpha}
           )
           (
              1
              +
              \varepsilon_{1\beta}
              \lambda_{1\beta}
           )
        }
   \left[
      \sum_{\nu=1}^{s} n_\nu
                       \frac{1-\varepsilon_{1\nu}\lambda_{1\nu}}
                            {1+\varepsilon_{1\nu}\lambda_{1\nu}}
   \right]^{-1}
   ,
$$
$$
   \eqno{\smash{(19)}}
$$
$$
   \theta_{\alpha}
   =
   n
   \theta
   \frac{1-\varepsilon_{1\alpha}\lambda_{1\alpha}}
        {1+\varepsilon_{1\alpha}\lambda_{1\alpha}}
   \left[
      \sum_{\beta=1}^{s} n_\beta
                         \frac{1-\varepsilon_{1\beta}\lambda_{1\beta}}
                              {1+\varepsilon_{1\beta}\lambda_{1\beta}}
   \right]^{-1}
   ,
$$
where $n=n_1+n_2+\ldots+n_s$ and $\theta$ is the mean temperature of
the mixture. Independent parameters $\lambda_{1\alpha}$ will be determined by
the condition of absence of energy redistribution between the components:
$$
   \sum_{\beta=1}^{s} \frac{
                              n_{\alpha\beta}
                              (
                                 1
                                 -
                                 \varepsilon_{1\alpha}
                                 \varepsilon_{1\beta}
                                 \lambda_{1\alpha}
                                 \lambda_{1\beta}
                              )
                              [
                                 (
                                    1
                                    -
                                    \varepsilon_{\alpha\beta}^2
                                 )
                                 (
                                    1
                                    -
                                    \lambda_{\alpha\beta}^2
                                 )
                              ]^{5/2}
                           }
                           {
                              (
                                 1
                                 +
                                 \varepsilon_{1\alpha}
                                 \lambda_{1\alpha}
                              )
                              (
                                 1
                                 +
                                 \varepsilon_{1\beta}
                                 \lambda_{1\beta}
                              )
                           }
                      \int\limits_{-\infty}^\infty
                          dx\,
                          \frac{x^2|1+x|^3}
                               {|1+x|^3-|1-x|^3}\,
                      \cdot
$$
$$
                      \cdot
                      \int\limits_0^1
                          dy\,\frac{
                          y^2
                          [
                             x^2
                             -
                             1
                             +
                             \varepsilon_{\alpha\beta}
                             y^2
                             (1+x)^2
                          ]
                          }{
                          [
                             (
                                1
                                +
                                \varepsilon_{\alpha\beta}
                             )
                             (
                                1
                                +
                                \varepsilon_{\alpha\beta}
                                \lambda_{\alpha\beta}
                             )
                             x^2 -
                            2
                             \lambda_{\alpha\beta}
                             (
                                1
                                -
                                \varepsilon_{\alpha\beta}^2
                             )
                             x
                             y^2
                             +
                             (
                                1
                                -
                                \varepsilon_{\alpha\beta}
                             )
                             (
                                1
                                -
                                \varepsilon_{\alpha\beta}
                                \lambda_{\alpha\beta}
                             )
                          ]^4
}
   =
   0
   .
   \eqno{(20)}
$$
Here $n_{\alpha\beta}$ are given by the number of single binary collisions
per unit volume of molecules with masses $m_\alpha$ and $m_\beta$. In solving
the system of equations (20) it is necessary to take into account
the equations (16), (18) expressing the dependent quantities
$\varepsilon_{\alpha\beta}$ and $\lambda_{\alpha\beta}$ in terms of
the independent ones. It is instructive to note also that the terms in (20)
with $\beta=\alpha$ are identically zero --- this is easy to see
changing the variable $x\to 1/x$.

The system (20) is too complicated for a general investigation. We are thus led
to restrict ourselves to the analysis of the two-component mixture carried out in
Section 2 which still gives some information about the main thermodynamical
characteristics of the mixture in more general case.

\section{Conclusion}

The general solution for distribution functions of a homogeneous
multicomponent mixture of free gases in thermodynamic
equilibrium found above predicts different partial temperatures of different
components. The ratios of partial temperatures to the mean temperature of
the mixture depend on molecular weights and concentrations of the components
--- see (13), (19). The effect of inequality of partial temperatures is
caused by the inequality of the collision probabilities for particles with velocities
${\bf v}_1 {\bf v}_2<0$ and ${\bf v}_1 {\bf v}_2>0$ --- see (6).
In the case of two-component mixture the maximal difference of
partial temperatures occurs at the ratio of molecular weights
$m_2/m_1\simeq 4.26$ and achieves approximately 15.5\% of minimal
temperature. As the ratio increases or decreases this difference decreases;
the character of decrease can be judged from figure 2 taking into account
(13). Given a homogeneous mixture of He and Ne atoms with equal
concentrations, at the mean temperature of mixture
$\theta\sim 300^\circ{\rm K}$ partial temperatures turn out to be
$\theta_{He}\sim 321^\circ{\rm K}$ and $\theta_{Ne}\sim 279^\circ{\rm K}$
which may well be verified experimentally.

In the case of multicomponent mixture the parameters $\lambda_{\alpha\beta}$
affecting the values of partial temperatures depend not only on molecular
weights but also on concentrations of the components --- see (20);
the lightest component temperature still exceeds the temperature of other
components and the mean temperature --- see (19).

The effect of inequality of partial temperatures disappears if the mixture due
to some reasons splits into separate regions with
domination of specific component in every region --- see (19); all
temperatures in such an event are practically equal to the mean temperature
of the whole system.

If different regions of the large system are filled with molecules of
different kinds and separated from each other by small potential barriers
(let us call such regions "cells") then under certain conditions an
interchange of light molecules not determined by the statistical dispersion
of velocities can appear between the "cells". Indeed, as the lightest
component temperature exceeds the mean temperature of the "cell", this
surplus maybe sufficient for getting over the barrier and penetrating into
the other "cell". There these light molecules can enter into reactions with
molecules of other kind resulting in particles lighter than
the molecules which fill this "cell". Transferring to these particles higher
temperature through collisions, this "cell" can push it out back in the first
"cell" and so forth. Penetration of foreign admixtures ("viruses") in "cells"
will cause redistribution of partial temperatures of its initial components.
Heavier admixtures will bring up the mean temperature of molecules originally
filling the "cell".

It is not improbable that in the case of multicomponent mixture,
selecting molecular weights and concentrations of components it is
possible to create a system with the property of enhanced conductivity at
not very low temperatures. This issue is based upon solution of the system of
equations (20) with simultaneous optimisation of the number of parameters
(involving concentrations), so it is a very complicated problem.
It is not clear also how stable the system constructed in such a way can be.

\bigskip\ \

{\bf Acknowledgements.} The author thanks P.\ Silaev for numerical assistance
and A.Sobolevsky for the manuscript preparation.

\baselineskip=16pt

\newpage
\section{References}
\begin{quote}

[1] Boltzmann L.
   Lectures on Gas Theory.-- Berkeley, Univ. of California Press, 1964.

[2] Leontovich M.\ A.
   Statistical Physics.-- Nauka,
   Moscow, 1944.

[3] Bogoliubov N.\ N.
   Problems of Dynamical Theory in Statistical Physics.-- Nauka,
   Moscow, 1946.

[4] Uhlenbeck G., Ford G.
   Lectures in Statisical Mechanics.-- AMS Press, Providence, 1963.

[5] Silin V.\ P.
   Introduction into Kinetic Theory of Gases.-- Nauka,
   Moscow, 1971.

[6] Landau L.\ D., Lifshitz E.\ M.
   Statistical Physics, Part 1.-- Nauka, Moscow, 1976.

[7] Lifshitz E.\ M., Pitaevskii L.\ P.
   Statistical Physics, Part 2.-- Nauka, Moscow, 1978.

[8] Balescu R.
   Equlibrium and Nonequlibrium Statistical Mechanics.-- John Wiley and Sons,
   New York, 1975.

[9] Kuni F.\ M.
   Statictical Physics and Thermodynamics.-- Nauka,
   Moscow, 1981.

[10] Klimontovich Yu.\ L.
   Statistical physics.-- Nauka,
   Moscow, 1982.

[11] Leontovich M.\ A.
Introduction into Thermodynamics. Statistical Physics.-- Nauka,
   Moscow, 1983.

[12] Kvasnikov I.\ A.
   Thermodynamics and Statistical Physics. Theory of Nonequlibrium Systems.--
   Moscow State University Press,
   Moscow, 1987.

[13] Loskutov Yu.\ M. Vestnik MGU, series 3: Physics, Astronomy.
   V.32 N.3 (1991) P.3--8.

[14] Feynman R., Leighton R., Sands M.
   The Feynman lectures on physics.
   Vol. 1.-- Addison-Wesley, London, 1963.

[15] Loskutov Yu.\ M. Fizicheskaya Mysl Rossii.
   N.3 (1995) P.1--9.
\end{quote}

\baselineskip=20pt

\newpage \

\vspace{1.0cm}

\begin{center}
\begin{figure}[htbp]
\centering
\includegraphics*[bbllx=0.19in,bblly=0.18in,bburx=4.92in,bbury=4.89in,scale=1.00]{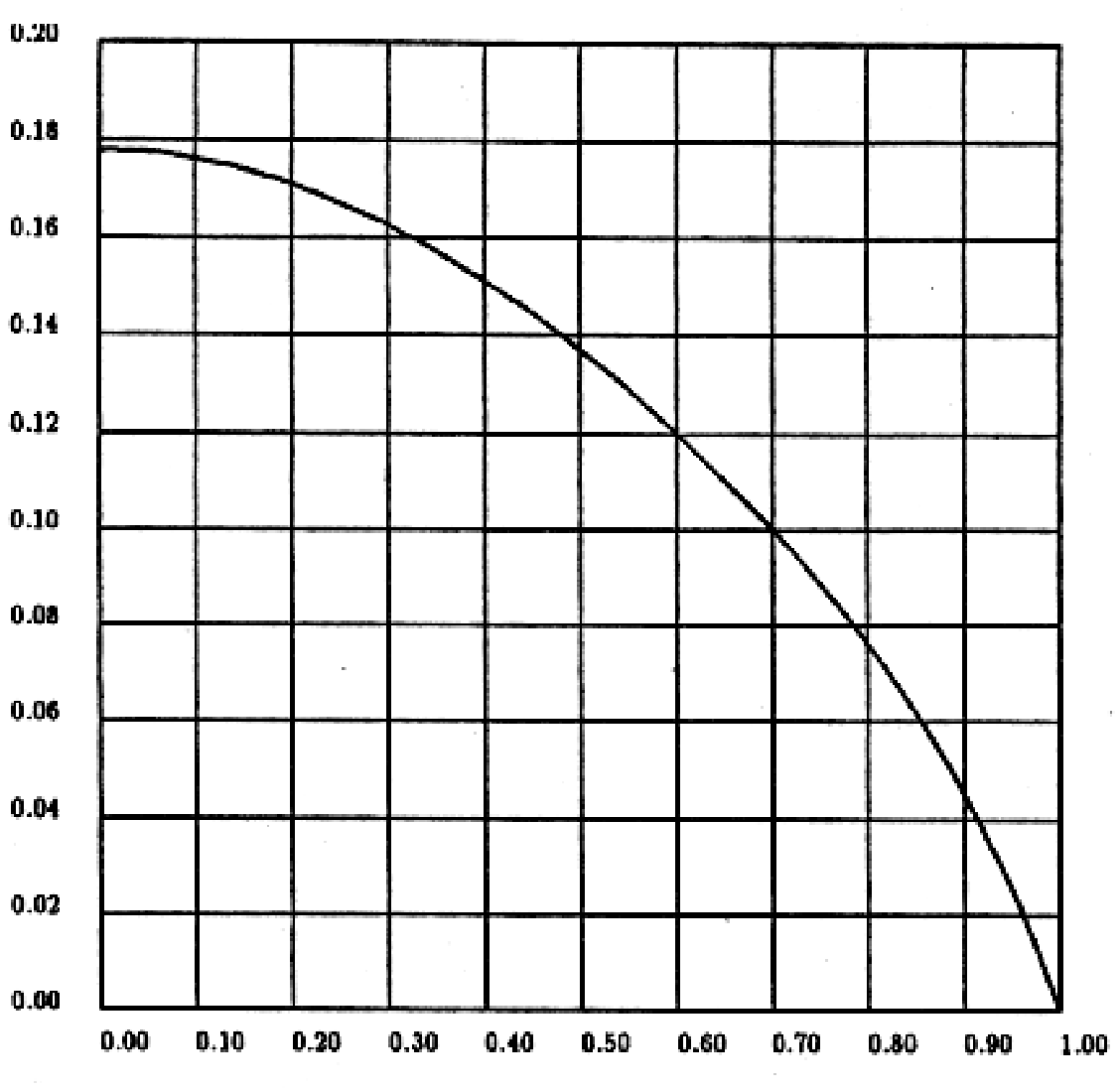}
\end{figure}

\end{center}

\begin{center}
Fig.1. Graphical dependence $\lambda $ versus $\varepsilon$.
\end{center}

\newpage

\vspace{1.0cm}

\begin{center}
\begin{figure}[htbp]
\centering
\includegraphics[scale=1.00]{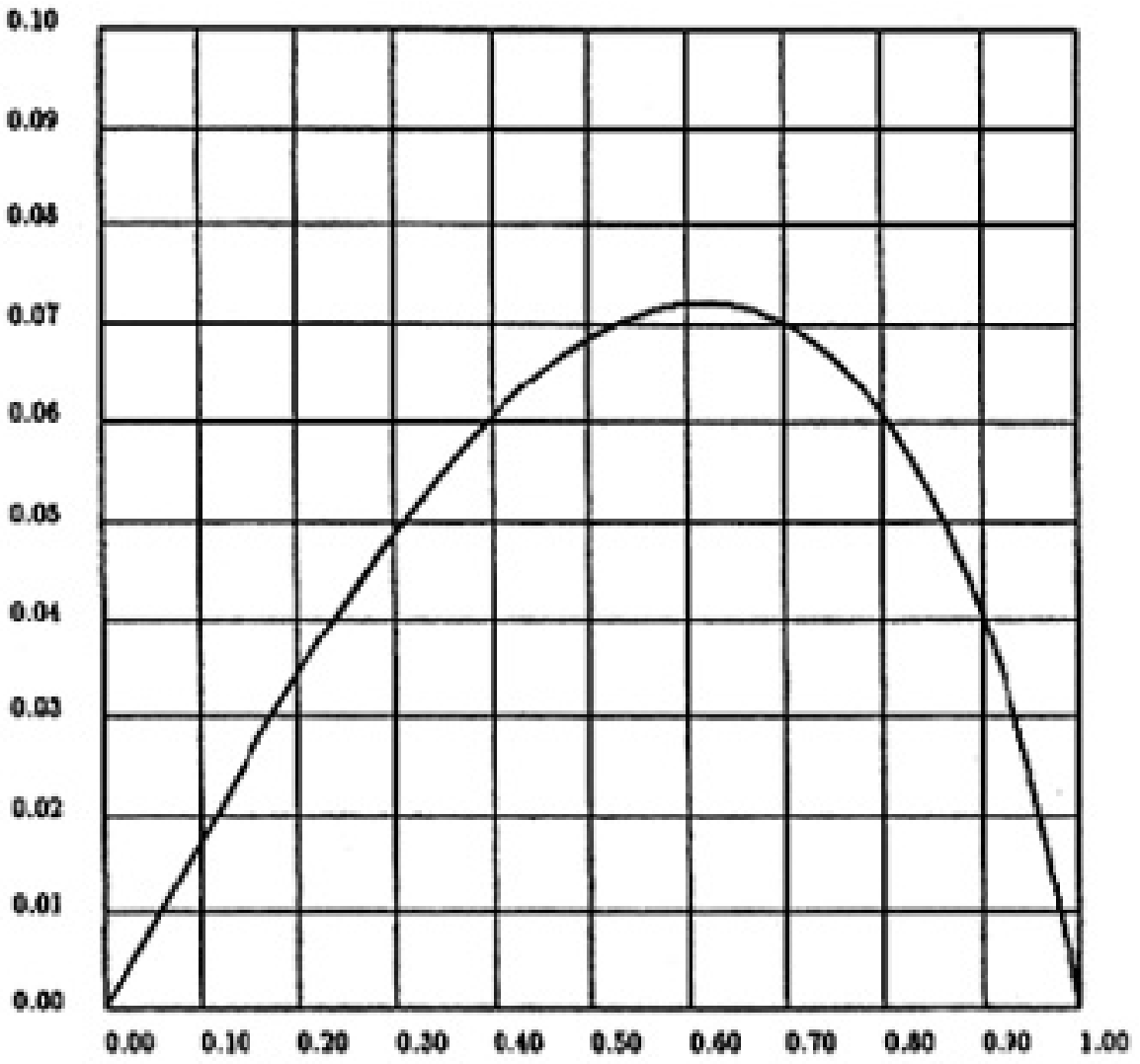}
\end{figure}

\end{center}

\begin{center}
Fig.2. Graphical dependence of $\varepsilon\lambda$ on $\varepsilon$
indirectly determining the ratio of partial temperatures.

\end{center}

\newpage
\vspace{1.0cm}

\begin{center}
Table 1. Numerical values of parameter $\lambda$ found from
equation (14) for different $\varepsilon$.
\end{center}

\begin{center}
\begin{figure}[htbp]
\centering
\includegraphics*[bbllx=0.19in,bblly=0.18in,bburx=4.54in,bbury=8.73in,scale=0.80]{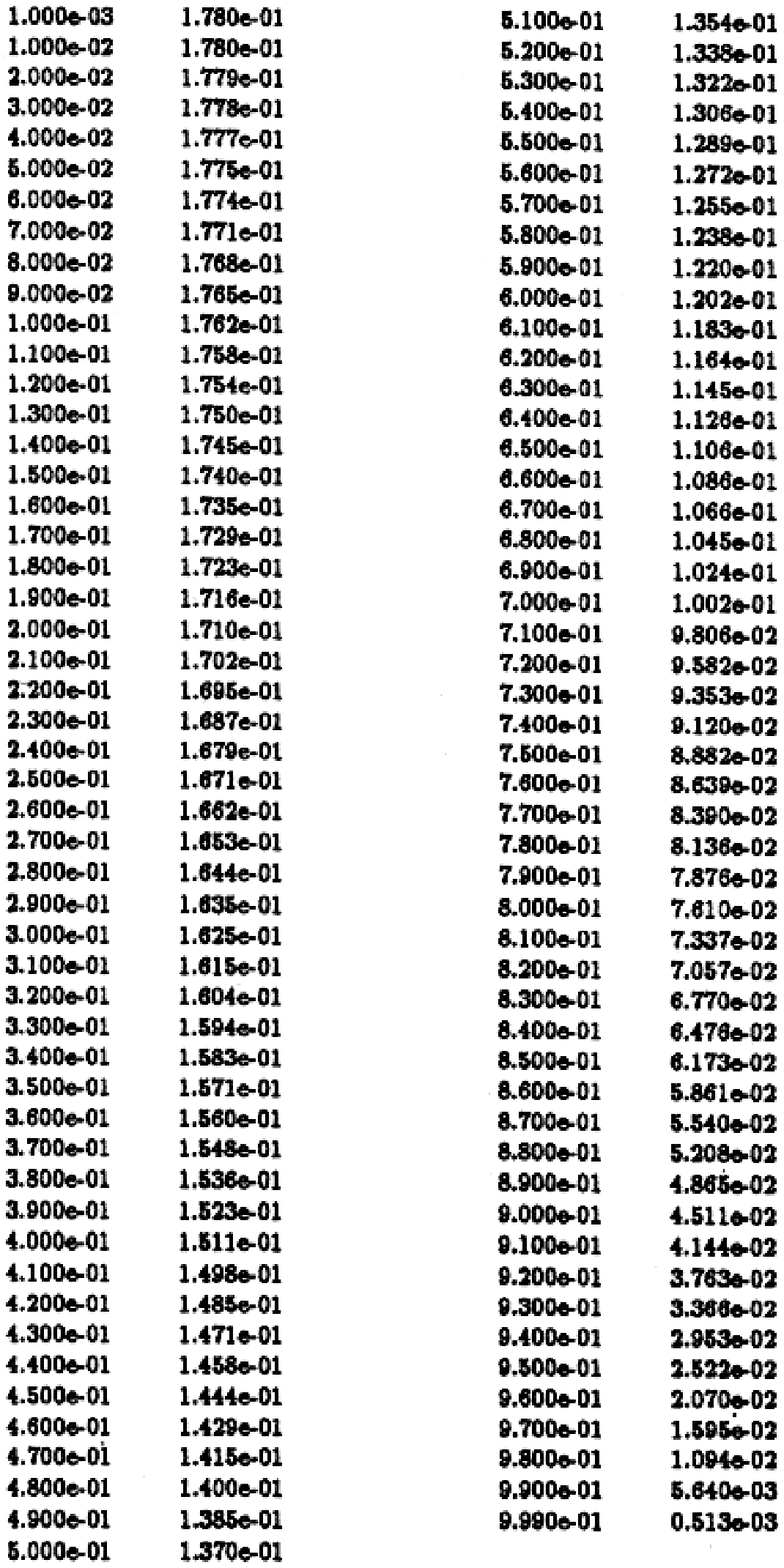}
\end{figure}

\end{center}

\end{document}